\documentclass[12pt]{article} %WS 10pt
\title{First-Order Field Equations in Spin 1/2 Form}  
\author{{\it Richard Shurtleff~}\thanks{affiliation and mailing 
address: Department of Applied Mathematics and Sciences, 
Wentworth Institute of Technology, 550 Huntington Avenue, 
Boston, MA, USA, ZIP 02115, telephone number: (617) 989-4338, fax 
number: (617) 989-4591 , e-mail address: shurtleffr@wit.edu}} 
\begin{document} 
          
\maketitle 

\begin{abstract} 
From one point of view in the quantum theory of fields, free quantum fields are uniquely determined, not by field equations, but by the transformations of the field and the annihilation and creation operators from which the field is constructed. One says that a free field equation merely records the fact that some field components are superfluous. Here, free field equations that are first order and covariant are derived so that the already determined field is one solution. The unknowns are the vector matrices $V^{\mu}$ that combine with the known gradient of the field, $\partial_{\mu} \psi,$ to make an invariant equation, $i V^{\mu} \partial_{\mu} \psi$ = $a \psi,$ where $a$ is a numerical constant. Thus these free field equations are direct consequences of the transformation properties of the annihilation and creation operators and the transformation properties of the field.

\vspace{0.5cm}
 
\vspace{0.5cm}

%03.65.Pm Relativistic wave equations;  03.70.+k Theory of quantized fields;   11.10.-z Field theory;  

\end{abstract}

\pagebreak

\section{Introduction} \label{intro}

First order covariant field equations for general spin have been studied going back almost to the equation for spin 1/2 found by Dirac. In much of this work, the field equations are solved for solution fields and the properties of particles are deduced. A few references are cited to give an idea of this process.\cite{FE1}-\cite{FE6} 

Before solving the equations, one must work out what matrices  $\beta^{\mu}$ give a relativistically invariant equation, 
\begin{equation} \label{beta} \beta^{\mu} \partial_{\mu} \psi = a \psi \quad, \end{equation} 
where $\psi$ is a field and $a$ is a constant. Many have solved this algebraic problem.\cite{Beta1}-\cite{Beta4} In this paper we derive the matrices $\beta^{\mu}$ again, but from a different point of view.

Because $a$ in (\ref{beta}) is a simple numerical constant, the matrices $\beta^{\mu}$ here are vector matrices $V^{\mu},$ 
\begin{equation} \label{Vbeta} \beta^{\mu} \equiv i V^{\mu} \quad, \end{equation} 
where a phase factor is included for convenience.

The process of starting with field equations and determining solution fields is just the opposite of the program followed here. Here the field has already been determined before the equations are derived. The field equations are derived so that the field is one solution.

One need not have a field equation to determine a field uniquely. Take as given the annihilation and creation operators for a particular particle with some known mass and spin. Consider the collection of fields that transform with a prescribed Lorentz representation. Then, as is well-known, one can select one field from the collection of all such fields by assuming that the field is a sum of coefficient functions times the known annihilation and creation operators. The essential ingredient is the Invariant Coefficient Hypothesis: The coefficient functions are invariant when the field transforms with a nonunitary Lorentz representation and the operators transform with a unitary Lorentz rep.

 Since the field is a sum of operators, there cannot be more independent field components than there are independent operators. We show that all fields with covariant first order field equations have more components than there are operators. Fields with some components dependent on others necessarily have `field equations' relating the components. Thus in a sense the field equations here record the fact that some field components are superfluous.\cite{Wb}

The fields here are not derived from the field equations, but are predetermined instead by the Invariant Coefficient Hypothesis. The field equations are derived from the known expressions for the field and the field is just one solution of those field equations. In this paper we obtain and discuss the subset of such field equations that are of first order and obviously covariant.

A list of some of what is not done in this paper may be helpful. We do not solve field equations. Spacial inversion, time reversal and charge conjugation are not considered here. The problem of interactions is not treated; all fields are free. The most general first order equations are not found; the numerical constant $a$ in (\ref{beta}) could instead be a matrix. While the vector matrices $V^{\mu}$ need not be hermitian or anti-hermitian, such possibilities are not excluded. Not all equations between field components can be put in the form (\ref{beta}). Equal time commutation relations are not imposed. 

The point is that the Invariant Coefficient Hypothesis is by itself powerful. Fields transform differently from the way the operators transform and yet the coefficient functions are invariant.  That spacetime symmetry gives field equations is interesting because it avoids the canonical methods that have evolved from classical physics. 

In Section \ref{Fields}, a standard process is presented in which fields are constructed as sums of annihilation and creation operators. Some well-known facts are recalled for convenience. In Section \ref{1CD}, it is useful to discuss the field equation for a field that transforms with a reducible rep of spin $(A,B)\oplus(C,D).$  The requirements on spins $A,B,C,D$ and particle spin $j$ are given in Section \ref{1CD} and the vector matrices are derived and discussed. In Section \ref{4CDs} the results of Section \ref{1CD} are used to obtain the first order covariant field equations for a field that transforms with a reducible rep of spin $(A,B)\oplus \sum_{i}(C_{i},D_{i}),$ combining $(A,B)$ with four irreducible reps $(C_{i},D_{i}).$ 
 The vector matrices $V^{\mu}$ found in Sec. \ref{4CDs} produce the equation
\begin{equation} \label{spinorFORM1} iV^{\mu}\partial_{\mu}\psi = a \psi \quad, \end{equation} 
where $a$ is an arbitrary constant. But the scale of $V^{\mu}$ is also arbitrary and, if $a$ is not zero $a \neq$ 0, we can force $a$ to be the mass $m$ of the particle by rescaling the vector matrices $V^{\mu}.$ The key to the method applied in Section \ref{4CDs} is to make the unknown vector matrices blockwise proportional to the matrices found in Section \ref{1CD}. The key works for fields with general spin combinations.

\section{Fields and Operators } \label{Fields}

In the quantum theory of fields, one may define a quantum field for a massive particle with mass $m$ and spin $j$ as a sum over the particle's annihilation and creation operators.\cite{W1}  The annihilation field $\psi^{(+)}(x)$ is a linear combination of annihilation operators, 
\begin{equation} \label{psiua1} \psi^{(+)}_{i} = \sum_{\sigma} u_{i \sigma} {\mathbf{a}}_{\sigma} \quad ,
 \end{equation} 
where ${\mathbf{a}}_{\sigma}(\overrightarrow{p})$ is the annihilation operator annihilating a state of the particle with spin component $\sigma,$ $\sigma \in$ $\{-j,-j+1,\ldots,j-1,j\},$ and $j$ is the spin of the particle. The momenta are determined by the three dimensional space $\overrightarrow{p}$ and the particle has mass $m.$ The time component of the momentum $p^{4}$ is determined by $p^{\mu}p_{\mu}$ = $m^{2}$ and $p^{4} >$ 0.

A creation field $\psi^{(-)}$ is a linear combination of creation operators,  
\begin{equation} \label{psiua2} \psi^{(-)}_{i} = \sum_{\sigma} v_{i \sigma} {\mathbf{b}}^{\dagger}_{\sigma} \quad ,
 \end{equation} 
where the creation operators ${\mathbf{b}}^{\dagger}_{\sigma}(\overrightarrow{p})$ may or may not be the adjoints of the annihilation operators ${\mathbf{a}}_{\sigma}$ in (\ref{psiua1}). The mass and spin of the particle is the same for both annihilation and creation operators, by assumption. Often the creation operators create a physical state for an antiparticle which can be considered a different particle.

Spacetime coordinates $x$ and the momenta $p$ are often suppressed. Each transforms with the usual Poincar\'{e} differential rep of Lorentz transformations with translations. For a combination $\Lambda$ of rotations and boosts followed by a translation through $\epsilon,$ one has $\psi(x) \rightarrow$ $\psi(\Lambda x + \epsilon) $ and ${\mathbf{a}}(\overrightarrow{p}) \rightarrow$ $\exp{(i\Lambda p \cdot \epsilon)}$ $  D^{(j)}[W^{(-1)}(\Lambda,p)]{\mathbf{a}}(\overrightarrow{p}_{\Lambda}),$ where $D^{(j)}[W(\Lambda,p)]$ represents the Wigner rotation corresponding to $\Lambda.$

The coefficient functions  $u_{i \sigma}(x,\overrightarrow{p})$ are functions of the spacetime coordinates $x$ that form the domain of the field and of the momenta $p$ that make up the domain of the operators. The behavior of $\psi(x)$ and ${\mathbf{a}}(\overrightarrow{p})$ under Poincar\'{e} transformations force  $u_{i \sigma}(x,p)$ to be plane waves, proportional to $\exp{(\pm i p_{\mu}x^{\mu})}$ so the sums (\ref{psiua1}) and (\ref{psiua2}) are often called `plane-wave expansions.' Thus each component of the field $\psi_{i}(x)$ satisfies the Klein-Gordon equation,
\begin{equation} \label{KG} \eta^{\mu \nu}\partial_{\mu}\partial_{\nu}\psi = m^{2} \psi \quad, \end{equation} 
where $\eta$ is the metric, $\eta$ = diag$\{1,1,1,-1\}.$ It is important to note that the field satisfies this second order Klein-Gordon equation because of the transformations of coordinates $x$ and momenta $p.$ This has nothing to do with the field equations obtained by component-counting.

A field, as a multicomponent object, is assumed here to transform with a non-unitary finite dimensional representation of the Lorentz group. Fields that transform via a Poincar\'{e} rep, which includes translations, will be considered elsewhere.

Consider the annihilation field, (\ref{psiua1}). The annihilation field $\psi^{(+)}$ transforms by the representation generated by angular momentum matrices $J.$ The annihilation operators ${\mathbf{a}}_{\sigma}$ transform by a spin $j$ representation generated by angular momentum matrices $J^{(j)}.$ The coefficient functions $u_{i \sigma}$ are invariant under these transformations, the Invariant Coefficient Hypothesis, which makes demands on the $u_{i \sigma}$ when they are sandwiched between the two different kinds of transformations. 

The essential condition that follows from the Invariant Coefficient Hypothesis is that the unitary representation transforming the operators and the nonunitary representation transforming the field must agree on the Wigner little group. For massive particles the Wigner little group is the group of rotations. It happens that the nonunitary representation transforming the field is unitary when restricted to rotations. 

Let $u^{0}_{i \sigma}$ be the coefficient functions in the rest frame, $\overrightarrow{p}$ = 0. (Recall the coefficient functions do depend on momentum, but that dependence is often suppressed in the notation.) The creation operators ${\mathbf{b}}^{\dagger}_{\sigma}$ transform differently from the annihilation operators ${\mathbf{a}}_{\sigma},$ while the fields $\psi^{(+)}$ and $\psi^{(-)}$ transform the same way. The fields  undergo a rotation generated by $\overrightarrow{J}.$ For the operators ${\mathbf{a}}_{\sigma}$ and ${\mathbf{b}}^{\dagger}_{\sigma},$ the rotation is generated by $\overrightarrow{J}^{(j)}$ and $-\overrightarrow{J}^{(j)\ast},$ respectively. Then the agreement of the two representations restricted to rotations requires that the generators $\overrightarrow{J}$ and $\overrightarrow{J}^{(j)}$ be compatible, see \cite{W1},
\begin{equation} \label{CG1} \overrightarrow{J}_{i\bar{i}}u^{0}_{\bar{i} \sigma} = u^{0}_{i \bar{\sigma}} \overrightarrow{J}^{(j)}_{\bar{\sigma} \sigma}  \quad,
 \end{equation}  
and
\begin{equation} \label{CGv1} \overrightarrow{J}_{i\bar{i}}v^{0}_{\bar{i} \sigma} = -v^{0}_{i \bar{\sigma}} \overrightarrow{J}^{(j)\ast}_{\bar{\sigma} \sigma}  \quad,
 \end{equation}
where $\overrightarrow{J}$ = $\{J^{x},J^{y},J^{z}\}$ = $\{J^{23},J^{31},J^{12}\}$ and the asterisk denotes complex conjugation. 

For the case considered in Section \ref{1CD}, $\psi$ is a field that transforms with a Lorentz rep of spin $(A,B)\oplus (C,D).$ The number $h$ of components of $\psi$ is then 
\begin{equation} \label{Order} h = (2A+1)(2B+1)+(2C+1)(2D+1) \quad .
 \end{equation}
With a similarity transformation if needed to obtain block-diagonal form, the $h \times h$ angular momentum matrices $J^{\mu \nu}$ are given by
\begin{equation} \label{blockFORM1} J^{\mu \nu} = \pmatrix{J_{(AB)}^{\mu \nu} && 0 \cr 0 && J_{(CD)}^{\mu \nu}} \quad .
 \end{equation}
Accordingly, the block form of the field separates the components that transform with the $(A,B)$ rep from those that transform via the $(C,D)$ rep,
\begin{equation} \label{blockFORM2} \psi^{(+)} = \pmatrix{\psi^{(+)}_{(AB)} \cr  \psi^{(+)}_{(CD)}} \quad {\mathrm{and}} \quad u_{\sigma} = \pmatrix{u_{(AB)\sigma} \cr  u_{(CD)\sigma}} \quad ,
 \end{equation}
and similarly for $\psi^{(-)}$ and $v_{i \sigma}.$
It may be convenient to transcribe the field index $i$  to a double index, $i \rightarrow $ $ab$ or $cd,$ where $a$ is the $z$-component of spin, $a \in$ $\{-A, -A+1, \ldots, A\},$ and likewise for $b,c,d.$

Eqns. (\ref{CG1}) are well-known, although they are usually considered for a Lorentz irrep, not the reducible rep here.  The solutions are blockwise proportional to Clebsch-Gordon coefficients. The overall factor is chosen to be the conventional factor $(2m)^{-1/2}$  and  the $CD$-block to $AB$-block ratios  $r_{u}$ and $r_{v}$ remain arbitrary. Thus the coefficient functions $u^{0}_{i \sigma}$ are determined by (\ref{CG1}) to be, see Ref. \cite{W1},
\begin{equation} \label{CG2} u^{0}_{i \sigma} = \left( \frac{1}{2m}\right)^{(1/2)}\pmatrix{\langle Aa;Bb\mid j \sigma \rangle \cr r_{u} \langle Cc;Dd\mid j \sigma \rangle} \quad,
 \end{equation}
where $\langle Aa;Bb\mid j \sigma \rangle$ indicates the Clebsch-Gordon coefficients connecting spin $(A,B)$  with spin $j$ and the index $i$ on the left in (\ref{CG2}) is replaced  by double indices $ab$ and $cd$ on the right. For the coefficient functions $v^{0},$ one can show that, see \cite{W1},
\begin{equation} \label{CGv2} v^{0}_{i \sigma} = (-1)^{j+\sigma} \left( \frac{1}{2m}\right)^{(1/2)}\pmatrix{\langle Aa;Bb\mid j (-\sigma) \rangle \cr r_{v} \langle Cc;Dd\mid j (-\sigma) \rangle} \quad.
 \end{equation}
Thus the coefficient functions are determined in the rest frame, $\overrightarrow{p}$ = 0.

By applying appropriate known transformation matrices, one can write the coefficient functions $u_{i \sigma}(x,\overrightarrow{p})$ and $v_{i \sigma}(x,\overrightarrow{p})$ for a general reference frame with arbitrary allowed momenta. Thus the fields $\psi^{(+)}(x)$ and $\psi^{(-)}(x)$ are essentially unique, unique up to an overall scale factor and the interblock ratios $r_{u}$ and $r_{v}$.

%\pagebreak

\section{Spin AB With One CD  } \label{1CD}

The field equations found in this section begin as identities between the Clebsch-Gordon coefficients in (\ref{CG2}) and (\ref{CGv2}) for the rest frame coefficient functions $u^{0}_{i \sigma}$ and $v^{0}_{i \sigma}.$ The agent that mixes the Clebsch-Gordon coefficients is the vector matrix $V^{t}.$ Thus we investigate how the vector matrix $V^{t}$ happens to have the needed properties to yield those Clebsch-Gordon identities.

A reducible Lorentz rep with spin $(A,B)\oplus(C,D)$ has vector matrices when the spins satisfy the requirement that $\mid A-C\mid$ = $\mid B-D\mid$ = 1/2, see Ref. \cite{Beta4},  as is true for spin $(1/2,0)\oplus(0,1/2),$ the Dirac rep.  There can be up to four `cases' $n \in$ $\{1,2,3,4\},$ of such reps $(C_{(n)},D_{(n)})$  associated with a given rep $(A,B),$  
\begin{eqnarray} \label{Cases1}  (C_{1},D_{1})=(A-1/2,B-1/2) \quad (C_{2},D_{2})=(A-1/2,B+1/2) \quad \cr \cr(C_{3},D_{3})=(A+1/2,B-1/2) \quad (C_{4},D_{4})=(A+1/2,B+1/2)\quad,
 \end{eqnarray}
i.e. $\mid A-C_{(n)} \mid$ = $\mid B-D_{(n)} \mid$ = 1/2 and, of course, the spins $C_{(n)}$ and $D_{(n)}$ must be nonnegative. 

The compatibility of the particle spin $j$ with the spins $A,B,C,D$ of the Lorentz rep evokes up-to-five triangle inequalities,
\begin{eqnarray} \label{ABCDj}  j \in \{\mid B-A \mid,\mid B-A \mid+1 \ldots , A+B-1,A+B\} \quad \cr \cr j \in \{\mid D_{(n)}-C_{(n)} \mid,\mid D_{(n)}-C_{(n)} \mid+1 \ldots , C_{(n)}+D_{(n)}-1,C_{(n)}+D_{(n)}\} \quad,
 \end{eqnarray} 
for the up-to-four cases of $(C_{(n)},D_{(n)})$ in (\ref{Cases1}). When the triangle inequality for one of the cases, say case $m,$ fails then the field components transforming with the $(C_{(m)},D_{(m)})$ rep vanish and can be dropped since the field equations for null components are trivial. 

We start with a field that transforms with a Lorentz rep of spin $(A,B)\oplus (C_{(n)},D_{(n)})$ for just one of the cases (\ref{Cases1}). Thus, with a similarity transformation if needed to obtain block-diagonal form, the angular momentum matrices $J^{\mu \nu}$ are in the form of (\ref{blockFORM1}) and the field $\psi_{i}$ is in the block form (\ref{blockFORM2}). The vector matrices $V^{\mu}$ are given by
\begin{equation} \label{blockFORM1V} V^{\mu } = \pmatrix{0&& V_{(AB;CD)}^{\mu} \cr V_{(CD;AB)}^{\mu}&&0} \quad.
 \end{equation}
There are no diagonal blocks because there are no vector matrices for a Lorentz irrep.\cite{Beta4}

The transformation property defining a vector matrix $V^{\mu}$ is
\begin{equation} \label{Comm2} D_{i \bar{i}}(\Lambda) V^{\mu}_{\bar{i}\bar{j}}D^{-1}_{ \bar{j}j}(\Lambda) = \Lambda^{\mu}_{\nu} V^{\nu}_{ij}  \quad,
\end{equation}
where $D(\Lambda)$ is the matrix generated by the $J^{\mu \nu}$ matrices in (\ref{blockFORM1}) representing the Lorentz transformation $\Lambda.$ 
Consider applying (\ref{Comm2}) to an infinitesimal Lorentz transformation $\Lambda^{\mu}_{\nu}$ = $\delta^{\mu}_{\nu}$ + $\omega^{\mu}_{\nu}$ and $D(\Lambda)$ = $ ({\mathbf{1}}-\frac{i}{2} \omega_{\alpha \beta} J^{\alpha \beta}),$ where $\omega_{\mu \nu}$ is antisymmetric and infinitesimal and ${\mathbf{1}}$ is the unit matrix. Simplifying the resulting equation,  one finds the commutation rules obeyed by angular momentum and vector matrices are
\begin{equation} \label{Comm1} i[V^{\mu},J^{\alpha \beta}]  = \eta^{\mu \alpha} V^{\beta} - \eta^{\mu \beta} V^{\alpha} \quad,
 \end{equation}
where $[V,J] \equiv$ $VJ-JV$ and $\eta$ = diag$\{+1,+1,+1,-1\}$ is the spacetime metric. 

The commutation rules (\ref{Comm1}) are obviously homogeneous in $V$. Less obviously, the rules are blockwise homogeneous and there is a free scale factor $N^{(AB;CD)}_{(n)}$ for the upper right block and an independent free parameter $N^{(CD;AB)}_{(n)}$ for the lower left block of the $V^{\mu}$ matrices (\ref{blockFORM1V}).

 By (\ref{Comm1}), $V^{t}_{(nu)}$ commutes with $\overrightarrow{J}$ = $\{J^{23},J^{31},J^{12}\}.$ Multiplying (\ref{CG1}) by the matrix $V^{t}_{(nu)}$ yields 
$$ (V^{t}_{(nu)})_{i\bar{i}} \overrightarrow{J}_{\bar{i}\bar{j}}u^{0}_{\bar{j} \sigma} = (V^{t}_{(nu)})_{i\bar{i}} u^{0}_{\bar{i} \bar{\sigma}} \overrightarrow{J}^{(j)}_{\bar{\sigma} \sigma}  \quad. $$ 
 By the commutation rule, interchange $V^{t}_{(nu)}$ and $ \overrightarrow{J}$ on the left, giving
\begin{equation} \label{CG3}  \overrightarrow{J}_{i\bar{i}} [(V^{t}_{(nu)})_{\bar{i}\bar{j}} u^{0}_{\bar{j} \sigma}] = [(V^{t}_{(nu)})_{i\bar{i}} u^{0}_{\bar{i} \bar{\sigma}}] \overrightarrow{J}^{(j)}_{\bar{\sigma} \sigma}  \quad.
 \end{equation}
Again we have the equations (\ref{CG1}); this time with $u^{0}_{ \sigma} \rightarrow$  $V^{t}_{(nu)} u^{0}_{ \sigma}.$ Both $V^{t}_{(nu)} u^{0}_{ \sigma}$ and $u^{0}_{ \sigma}$ satisfy the same equation and they are closely related. 

This answers the question posed earlier. The matrix $V^{t}$ is special because $V^{t}$ commutes with $ \overrightarrow{J}$ which generates the group of rotations which is the Wigner little group of a massive particle. 

We show that $V^{t}_{(nu)} u^{0}_{ \sigma}$ and $u^{0}_{ \sigma}$ are proportional when $V^{t}_{(nu)}$ has suitable block scale factors. In the Appendix we use a formula for the Clebsch-Gordon coefficients with $\sigma$ = $j$ to find the scale factors $N^{(AB;CD)}_{(nu)}$ and $N^{(CD;AB)}_{(nu)}$ so that $V^{t}_{(nu)} u^{0}_{ \sigma}$ and $u^{0}_{ \sigma}$ are proportional,
\begin{equation} \label{Field1} V^{t}_{(nu)} u^{0}_{\sigma} = s_{u} u^{0}_{\sigma}  \quad,
 \end{equation}
where $s_{u}$ is an arbitrary constant factor representing the freedom to scale $V^{t}.$ The equality holds for all allowed $\sigma.$  The resulting values of $N^{(AB;CD)}_{(nu)}$ and $N^{(CD;AB)}_{(nu)}$ are collected in Table 1 with the stipulation that one replaces the $r$ and $s$ in Table 1 by $r_{u}$ and $s_{u}$. 

Similarly operating on (\ref{CGv1}) by multiplying with a possibly different vector matrix $V^{t}_{(nv)},$ one finds that 
 \begin{equation} \label{Fieldv1} V^{t}_{(nv)} v^{0}_{\sigma} = s_{v} v^{0}_{\sigma}  \quad,
 \end{equation}
for a new arbitrary constant $s_{v}$ and for all allowed $\sigma.$  Formulas for $N^{(AB;CD)}_{(nv)}$ and $N^{(CD;AB)}_{(nv)}$ are collected in Table 1; replace $r$ and $s$ in Table 1 by $r_{v}$ and $s_{v}$.

\begin{table} [ht]
\caption{{\it{Scale Factors For Vector Matrix Blocks.}} 
These values give $V^{\mu}_{(nu)}$ and $V^{\mu}_{(nv)}$when used in the vector matrices formulas of Ref. \cite{Beta4} where $N^{(AB;CD)}_{(n)}$
	and $N^{(CD;AB)}_{(n)}$ are denoted $t^{12}_{AB}$  and $t^{21}_{CD},$ respectively. Parameters $r_{u}$ and $r_{v}$ are the ratios of the $CD$ to $AB$ coefficients in (\ref{CG1}) and (\ref{CGv1}), while $s_{u}$ and $s_{v}$ are the eigenvalues in (\ref{Field1}) and (\ref{Fieldv1}). 
\vspace{0.5cm}}
\begin{tabular}{|ccccc|}
	\hline
\textbf{\em $n$}
	& $\frac{r}{s}N^{(AB;CD)}_{(n)}$
	& {$\frac{1}{rs}N^{(CD;AB)}_{(n)}$} &  $C_{(n)}$ &  {$D_{(n)}$}
	 \\ \hline \hline
1   & $\frac{2\sqrt{AB}}{\sqrt{(A+B)(A+B+1)-j(j+1)}}$ & $\frac{1}{\sqrt{(A+B)(A+B+1)-j(j+1)}}$    & $A-1/2$  & $B-1/2$             \\
2   &  $\frac{-\sqrt{2A}}{\sqrt{j(j+1)-(B-A)(B-A+1)}}$ & $\frac{\sqrt{2B+1}}{\sqrt{j(j+1)-(B-A)(B-A+1)}}$  & $A-1/2$  & $B+1/2$  \\
3   &  $\frac{\sqrt{2B}}{\sqrt{j(j+1)-(A-B)(A-B+1)}}$ & $\frac{-\sqrt{2A+1}}{\sqrt{j(j+1)-(A-B)(A-B+1)}}$  & $A+1/2$  & $B-1/2$  \\
4   &   $\frac{1}{\sqrt{(A+B+1)(A+B+2)-j(j+1)}}$  & $\frac{\sqrt{(2A+1)(2B+1)}}{\sqrt{(A+B+1)(A+B+2)-j(j+1)}}$   & $A+1/2$  & $B+1/2$    \\ \hline
\end{tabular}
\end{table}

%\pagebreak

One may recognize (\ref{Field1}) and (\ref{Fieldv1}) as field equations in the rest frame. To transform these equations to a general reference frame, first multiply by the mass $m.$ Then multiply by the matrix $D(L(p))$ representing a special transformation $L(p)$ that takes the rest frame momentum $p^{\mu}_{0}$ = $\{0,0,0,m\}$ to a generic momentum $p^{\mu}$ = $L^{\mu}_{\nu} p^{\nu}_{0}.$ By the transformation properties of vector matrices (\ref{Comm2}) one finds that
$$V^{t}_{(nu)} u^{0}_{\sigma} = s_{u} u^{0}_{\sigma} \quad $$
$$m D(L(p))V^{t}_{(n)} D(L(p))^{-1} D(L(p)) u^{0}_{ \sigma}= m s_{u} D(L(p))u^{0}_{ \sigma} $$
$$m L^{t}_{\nu}V^{\nu}_{(n)}  u_{\sigma} = m s_{u} u_{\sigma}$$
\begin{equation} \label{Field2} -p_{\nu}V^{\nu}_{(nu)} u_{\sigma} = s_{u} m u_{\sigma} \quad , \end{equation}
where $u$ = $ D(L(p)) u^{0}$ is the coefficient function for momentum $p^{\mu}.$ 

For the creation field, one finds
\begin{equation} \label{Fieldv2} -p_{\mu}V^{\mu}_{(nv)} v_{\sigma} = s_{v} m v_{\sigma} \quad , \end{equation}
where $v$ = $ D(L(p)) v^{0}$ is the coefficient function for momentum $p^{\mu}.$

The dependence of $u_{i\sigma}$ and $v_{i\sigma}$ on coordinates $x^{\mu}$ and on momentum $p^{\mu}$ is suppressed in this paper, hidden in the subscripts $i$ and $\sigma,$ and sometimes as in (\ref{Field2}) and (\ref{Fieldv2}) the index $i$ is suppressed. As noted previously, the transformations of coordinates $x^{\mu}$ and momenta $p^{\mu}$ includes translations, forcing the coefficient functions to be plane waves, see Ref. \cite{W1},
\begin{equation} \label{uxp}u_{i\sigma} = u_{i\sigma}(x^{\mu},\overrightarrow{p}) = e^{ip_{\mu}x^{\mu}} u_{i\sigma}(0,\overrightarrow{p}) \quad , \end{equation}
\begin{equation} \label{vxp}v_{i\sigma} = v_{i\sigma}(x^{\mu},\overrightarrow{p}) = e^{-ip_{\mu}x^{\mu}} v_{i\sigma}(0,\overrightarrow{p}) \quad . \end{equation}
Note the sign difference in the phase factors. 

Hence, multiplication of $u_{i\sigma}$ by a component of momentum $p_{\mu}$ is equivalent to partial differentiation with respect to the coordinate $x^{\mu},$ i.e $p_{\mu} \rightarrow$ $-i \partial_{\mu}.$ And multiplication of $v_{i\sigma}$ by a component of momentum $p_{\mu}$ is equivalent to partial differentiation with respect to the coordinate $x^{\mu},$ i.e $p_{\mu} \rightarrow$ $+i \partial_{\mu}.$ Thus, by (\ref{psiua1}) and (\ref{Field2}), we have obtained the field equations  
\begin{equation} \label{Field3} i V^{\mu}_{(nu)} \partial_{\mu} \psi^{(+)} = s_{u} m \psi^{(+)}  \quad.
 \end{equation}
And by (\ref{psiua2}) and (\ref{Fieldv2}), we have obtained the field equations  
\begin{equation} \label{Fieldv3} i V^{\mu}_{(nv)} \partial_{\mu} \psi^{(-)} = -s_{v} m \psi^{(-)}  \quad,
 \end{equation}
where the sign difference $-s_{v}$ versus $+s_{u}$ reflects the sign difference in (\ref{uxp}) and (\ref{vxp}).

The annihilation and creation fields $\psi^{(+)}$ and $\psi^{(-)},$ must transform by the same Lorentz rep to combine to form a field $\psi,$
\begin{equation} \label{psi} \psi_{i} = \kappa \psi^{(+)}_{i} + \lambda \psi^{(-)}_{i} \quad ,
 \end{equation} 
where $\kappa$ and $\lambda$ are arbitrary constants. However for the gradient of $\psi$ to be part of an equation, it is necessary to constrain a couple of arbitrary constants in $\psi^{(+)}$ and $\psi^{(-)}$ because the phase $\exp{(ip_{\mu}x^{\mu})}$ in the annihilation plane waves clashes with the phase $\exp{(-ip_{\mu}x^{\mu})}$ in the creation plane waves. A similar situation occurs with the Dirac equation.\cite{W1}

Now we impose conditions on the vector matrices and the fields so that $\psi^{(+)}$ and $\psi^{(-)}$ obey the same field equation. First we require the vector matrices to be the same,
 \begin{equation} \label{equalV}  V^{\mu}_{(nu)} =  V^{\mu}_{(nv)} \quad.
 \end{equation}
The vector matrices are the same when the block scale factors of $V^{\mu}_{(nu)}$ and $V^{\mu}_{(nv)}$ are equal, $N_u$ = $N_{v}.$ By Table 1, we have
 \begin{equation} \label{equalV2}  \frac{s_{u}}{r_{u}} =  \frac{s_{v}}{r_{v}} \quad {\mathrm{and}} \quad r_{u}s_{u} =  r_{v}s_{v}  \quad.
\end{equation}
Second, the sign difference in (\ref{Field3}) and (\ref{Fieldv3}) must be dealt with. To do this we constrain the arbitrary constants $r_{v}$ and $r_{u}$ which exist because (\ref{CG1}) and (\ref{CGv1}) are blockwise homogeneous in the coefficient functions $u_{i\sigma}$ and $v_{i\sigma}.$ Assume $r_{v}$ = $-r_{u}.$ Then, by (\ref{equalV2}) one finds that
 \begin{equation} \label{Sign} r_{v} =  -r_{u} \quad {\mathrm{implies}} \quad s_{v} =  -s_{u}  \quad,
\end{equation}
which makes the signs agree in (\ref{Field3}) and (\ref{Fieldv3}). 

For the Dirac field the ratios $r_{u}$ and $r_{v}$ occur as the ratio of spin $(0,1/2)$ to spin $(1/2,0)$ components and these have opposite parity under spacial inversion. Thus, for the Dirac field the assumption $r_{v}$ = $-r_{u}$ can be interpreted as a statement about `parity.'\cite{W3} But spins $(A,B)$ and $(C_{(n)},D_{(n)})$ do not in general have opposite parity. For example, consider case 2 for spin $(A,B)$ = $(1/2,1/2)$ combined with $(C_{(2)},D_{(2)})$ = $(0,1).$ Spin $(1/2,1/2)$ is itself parity invariant, but the spin with the opposite parity of $(0,1)$ is $(1,0).$ The interblock ratios still need to be adjusted even though $(1/2,1/2)\oplus(0,1)$ is not simply behaved under spacial inversion (parity). Therefore, adjusting $r_{u}$ and $r_{v}$ to allow compatible differentiation of the plane waves $\exp{(ip_{\mu}x^{\mu})}$ and $\exp{(-ip_{\mu}x^{\mu})}$ is not related to space inversion, in general.  

 To simplify the parameters further, one may absorb the parameters $s_{u}$ and $s_{v}$ into the scale factors, $N^{AB;CD}_{(n)}$ = $N^{AB;CD}_{(nu)}/s_{u},$ etc. The parameter $s_{u}$ cannot be zero because that would make the vector matrices vanish. Therefore, we can have 
 \begin{equation} \label{JUSTs}  s_{u} =  -s_{v} = 1  \quad.
\end{equation}
The vector matrices are now equal (\ref{equalV2}), the signs adjusted (\ref{Sign}) and the vectors rescaled (\ref{JUSTs}) to simplify the field equations.

By (\ref{Field3}), (\ref{Fieldv3}) and (\ref{JUSTs}), the field $\psi$ = $\kappa \psi^{(+)} + \lambda \psi^{(-)} $ in (\ref{psi}) satisfies the field equation
\begin{equation} \label{Field4} i V^{\mu}_{(n)} \partial_{\mu} \psi =  m \psi  \quad,
 \end{equation}
where the vector matrices $V^{\mu}_{(n)}$ are calculated with the formulas of Ref. \cite{Beta4} and the block scale factors in Table 1 with $s$ = 1. The equations (\ref{Field4}) apply to a field $\psi$ transforming with a Lorentz rep with spin $(A,B)\oplus(C_{(n)},D_{(n)}),$ for a single case $n.$

{\it{Remarks.}} When spin $(A,B)$ is scalar $(0,0),$ only case $n = 4$ gives nonnegative $C_{(n)}$ and $D_{(n)},$ and the result (\ref{Field4}) is appropriate. And for other spins $(A,B),$ one may be interested in just one case. The Dirac formalism involves spin $(0,1/2)\oplus(1/2,0)$ which is case 3, thereby leaving out case 4 which also works with spin $(0,1/2).$ Thus, the field equations found in this section for Lorentz reps with spin $(A,B)\oplus(C_{(n)},D_{(n)})$ for a single case $n$ can be appropriate for some applications.

Other situations require more complicated spin combinations. Spin $(1/2,1/2)$ $\oplus(0,1)\oplus(1,0),$ for example, has parity structure that is lost if either spin $(0,1)$ or spin $(1,0)$ field components are left out. And an irrep with spin $(C_{(n)},D_{(n)})$ can be associated with some other irrep with spin $(E,F)$ that satisfies the test for vector matrices $\mid E-C_{(n)} \mid$ = $\mid F-D_{(n)} \mid$ = 1/2. In the next section a method of handling more complicated spin combinations is illustrated by treating an example, the Lorentz rep with spin $(A,B)\oplus \sum_{n}(C_{(n)},D_{(n)}),$ summing over all four allowed cases $n$ in (\ref{Cases1}).

\section{Spin AB With Four CDs  } \label{4CDs}

As an example of a fairly complicated spin structure, consider the reducible Lorentz rep with spin $(A,B)\oplus \sum_{n}(C_{(n)},D_{(n)}),$ obtained by combining the irreducible rep with spin $(A,B)$ with the irreps for all four cases $(C_{(n)},D_{(n)})$ in (\ref{Cases1}). Assume that neither $A$ nor $B$ is zero so that all of the values of $C_{(n)}$ and $D_{(n)}$ in (\ref{Cases1}) are nonnegative. 

The angular momentum matrices can be written in a $5 \times 5$ block form, 
\begin{equation} \label{blockFORM3} J^{\mu \nu} = \pmatrix{J_{(AB)}^{\mu \nu} && 0 && 0 && 0 && 0 \cr 0 && J_{C_{1}D_{1}}^{\mu \nu} && 0 && 0 && 0 \cr 0 && 0 && J_{C_{2}D_{2}}^{\mu \nu} && 0 && 0  \cr 0 && 0 && 0 && J_{C_{3}D_{3}}^{\mu \nu} && 0  \cr 0 && 0 && 0 && 0 && J_{C_{4}D_{4}}^{\mu \nu}}   \quad .
 \end{equation}
The block form of the field $\psi$ is
\begin{equation} \label{blockFORM5} \psi = \pmatrix{\psi_{(AB)} \cr  \psi_{(C_{1}D_{1})}\cr  \psi_{(C_{2}D_{2})}\cr  \psi_{(C_{3}D_{3})}\cr  \psi_{(C_{4}D_{4})}} \quad .
\end{equation}

The spins of all blocks $V_{(C_{(n)}D_{(n)};C_{(m)}D_{(m)})}^{\mu}$ differ by integers and so fail to have the required difference of 1/2. The corresponding blocks vanish leaving the vector matrices with a peculiar block matrix form 
\begin{equation} \label{blockFORM4}  V^{\mu } = \pmatrix{0 && V_{(AB;C_{1}D_{1})}^{\mu} && V_{(AB;C_{2}D_{2})}^{\mu} &&V_{(AB;C_{3}D_{3})}^{\mu} &&V_{(AB;C_{4}D_{4})}^{\mu} \cr V_{(C_{1}D_{1};AB)}^{\mu}&&0&& 0 &&0 && 0 \cr V_{(C_{2}D_{2};AB)}^{\mu}&&0&& 0 &&0 && 0 \cr V_{(C_{3}D_{3};AB)}^{\mu}&&0&& 0 &&0 && 0 \cr V_{(C_{4}D_{4};AB)}^{\mu}&&0&& 0 &&0 && 0} \quad.
 \end{equation}
 For any case $n,$ spin $A$ differs from $C_{(n)}$ by $\pm 1/2$ so $C_{(n)}$ and $C_{(m)}$ differ by either 0 or $\pm 1.$ Similar reasoning applies to $D_{(n)}$ and $D_{(m)}$. Since $\mid C_{(n)} - C_{(m)} \mid \neq$ 1/2 and $\mid D_{(n)} - D_{(m)} \mid \neq$ 1/2, these spins fail the test for having vector matrices and the corresponding blocks vanish.

Each nonzero block, such as $V_{(AB;C_{1}D_{1})}^{\mu},$ of the vector matrices is determined within an arbitrary factor by the commutation rules (\ref{Comm1}), which are homogeneous in $V_{(AB;C_{1}D_{1})}^{\mu}.$ Thus, without loss of generality, one may define the arbitrary factor so that $V_{(AB,C_{1}D_{1})}^{\mu}$ is some other arbitrary factor, say ${a}_{1}/m,$  times $V_{(AB;CD;1)}^{\mu},$ which is a block of the vector matrix $V^{\mu}_{(1)}$ obtained in Sec. \ref{1CD}. Then, for the four cases $n,$ we have
\begin{equation} \label{Vblock1} V_{(AB,C_{(n)}D_{(n)})}^{\mu} = a_{n} m^{-1} V^{\mu}_{(AB,CD;n)} \quad {\mathrm{and}} \quad V_{(C_{(n)}D_{(n)},AB)}^{\mu} = b_{n} m^{-1} V^{\mu}_{(CD,AB;n)} \quad ,
 \end{equation}
where ${a}_{n}$ and ${b}_{n}$ are arbitrary, and $m$ is the mass of the particle. The mass $m$ cannot be zero because we are considering nonzero mass particles and have assumed that the annihilation and creation operators have rotations as the Wigner little group. The subscript `$AB,CD;n$' indicates the upper right block of $V^{\mu}_{(n)}$ and `$CD,AB;n$' indicates the lower left block of the matrix $V^{\mu}_{(n)}$ obtained in Sec. \ref{1CD}, i.e. the formulas of Ref. \cite{Beta4} using the parameters $N^{(AB;CD)}_{(n)}$ and $N^{(CD;AB)}_{(n)}$ in Table 1 with $s$ = 1. 

By (\ref{Vblock1}), the vector matrices (\ref{blockFORM4}) can now be written as follows,
\pagebreak
$$V^{\mu } =   \hspace{15cm} $$
\begin{equation} \label{Vblock2}  m^{-1}\pmatrix{0 && a_{1} V_{(AB,CD;1)}^{\mu} && a_{2}V_{(AB,CD;2)}^{\mu} && a_{3}V_{(AB,CD;3)}^{\mu} && a_{4}V_{(AB,CD;4)}^{\mu} \cr b_{1}V_{(CD,AB;1)}^{\mu}&&0&& 0 &&0 && 0 \cr b_{2}V_{(CD,AB;2)}^{\mu}&&0&& 0 &&0 && 0 \cr b_{3}V_{(CD,AB;3)}^{\mu}&&0&& 0 &&0 && 0 \cr b_{4}V_{(CD,AB;4)}^{\mu}&&0&& 0 &&0 && 0} 
 \end{equation}

The new notation allows us to use the following blockwise relations from (\ref{blockFORM1V}) and (\ref{Field4}) in Sec. \ref{1CD}, 
\begin{equation} \label{Field4a} iV_{(AB,CD;n)}^{\mu} \partial_{\mu} \psi_{(C_{(n)}D_{(n)})} = m \psi_{(AB)} \quad {\mathrm{and}} \quad iV_{(CD,AB;n)}^{\mu} \partial_{\mu} \psi_{(AB)} = m \psi_{(C_{(n)}D_{(n)})} \quad.
 \end{equation}
By (\ref{blockFORM5}), (\ref{Vblock2}), and (\ref{Field4a}) one has 
\begin{equation} \label{Field5}  i V^{\mu} \partial_{\mu} \psi = \frac{1}{m} \pmatrix{\sum_{n} a_{n} i V_{(AB,CD;n)}^{\mu} \partial_{\mu} \psi_{(C_{(n)}D_{(n)})} \cr  b_{1} iV_{(CD,AB;1)}^{\mu} \partial_{\mu} \psi_{(AB)} \cr  b_{2} iV_{(CD,AB;2)}^{\mu} \partial_{\mu} \psi_{(AB)}\cr  b_{3} iV_{(CD,AB;3)}^{\mu} \partial_{\mu} \psi_{(AB)}\cr  b_{4} iV_{(CD,AB;4)}^{\mu} \partial_{\mu} \psi_{(AB)} } =   \pmatrix{(\sum_{n} a_{n})  \psi_{(AB)} \cr  b_{1}  \psi_{(C_{(1)}D_{(1)})} \cr  b_{2} \psi_{(C_{(2)}D_{(2)})}\cr  b_{3}  \psi_{(C_{(3)}D_{(3)})} \cr  b_{4}  \psi_{(C_{(4)}D_{(4)})} }\quad.
 \end{equation}
The expression on the far right is proportional to $\psi$ in (\ref{blockFORM5}) with arbitrary factor $a$ when 
\begin{equation} \label{Field5a}  \sum_{n} a_{n}  = a \quad {\mathrm{and}} \quad b_{1} = b_{2} = b_{3} = b_{4} = a  \quad,
 \end{equation}
which is what we assume. By (\ref{Field5}) and (\ref{Field5a}), we have 
\begin{equation} \label{Field6}  i V^{\mu} \partial_{\mu} \psi =  a  \psi \quad.
 \end{equation}
The vector matrices $V^{\mu}$ are be calculated from (\ref{Vblock2}) and (\ref{Field5a}) using Table 1 in Sec. \ref{1CD} with $s$ = 1 and Ref. \cite{Beta4} to calculate $V_{(AB,CD;n)}^{\mu}$ and $V_{(CD,AB;n)}^{\mu}.$ Equation (\ref{Field6}) and the method used to obtain it are the central results of this paper. 

{\it{Remarks.}} Consider what happens when the constant $a$ vanishes, $a$ = 0. By (\ref{Vblock2}), (\ref{Field5}) and (\ref{Field5a}), one has
\begin{equation} \label{Field9}  iV^{\mu}_{0} \partial_{\mu} \psi = 0 \quad,
 \end{equation}
with vector matrices $V^{\mu }_{0}$ found to be 
\begin{equation} \label{Vblock3}  V^{\mu }_{0} = m^{-1}\pmatrix{0 && a_{1} V_{(AB,CD;1)}^{\mu} && a_{2}V_{(AB,CD;2)}^{\mu} && a_{3}V_{(AB,CD;3)}^{\mu} && a_{4}V_{(AB,CD;4)}^{\mu} \cr 0&&0&& 0 &&0 && 0 \cr 0&&0&& 0 &&0 && 0 \cr 0&&0&& 0 &&0 && 0 \cr 0&&0&& 0 &&0 && 0} \quad,
 \end{equation}
and with $\sum_{n} a_{n}$ = $a$ = 0. Immediately, one sees that the $(A,B)$ block of $\psi$ is not part of the field equations (\ref{Field9}). Only the $(C_{(n)},D_{(n)})$ blocks of $\psi$ are involved in these field equations.

One might worry that the choice $a$ = 0 violates the fact that each component of $\psi$ satisfies the Klein-Gordon equation (\ref{KG}). To investigate this apply (\ref{Field9}) twice,
\begin{equation} \label{Field10}  -\frac{1}{2}(V^{\mu}_{0}V^{\nu}_{0}+ V^{\nu}_{0}V^{\mu}_{0})\partial_{\mu} \partial_{\nu}\psi = 0 \quad.
 \end{equation}
This would violate the Klein-Gordon equation (\ref{KG}) if, with a similarity transformation if needed, one could show that 
\begin{equation} \label{KGok}  (V^{\mu}_{0}V^{\nu}_{0}+ V^{\nu}_{0}V^{\mu}_{0})_{ij} = 2 \eta^{\mu \nu} \delta_{ij} \quad,
 \end{equation}
for at least one value of $i.$ But, by the block form of the matrices $V^{\mu}_{0}$ in (\ref{Vblock3}), it is obvious that $V^{\mu}_{0}V^{\nu}_{0}$ = 0. Thus the choice $a$ = 0 does not contradict the Klein-Gordon equation. More basically, note that there are five Lorentz irreps in the spin $(A,B)\oplus \sum_{n}(C_{(n)},D_{(n)}).$ With more than two irreps addition occurs in the overall scale factor as in (\ref{Field5a}) and cancelations are possible. This contrasts with the two irrep spin $(A,B)\oplus (C,D)$ in Section \ref{1CD} where addition (\ref{Field5a}) does not occur and $a$ = 0 would make the vector matrices vanish.

As an example with $a$ = 0, consider a spin 1 massive particle field $\psi$ that transforms with a Lorentz rep of spin $(1/2,1/2)\oplus(0,1)\oplus(1,0),$ i.e. cases 2 and 3 with case 4 left out.  The field equation (\ref{Field9}) involves only the spin $(0,1)\oplus(1,0)$ parts of the field $\psi.$ These parts of $\psi$ transform like an antisymmetric tensor $F^{\mu \nu}$ and the field equation (\ref{Field9}) can be shown to be a set of homogeneous Maxwell equations for $F^{\mu \nu},$ see Problem 6. Thus the homogeneous Maxwell equations for an antisymmetric field $F^{\mu \nu}$ can be put in the form (\ref{Field9}), but only if $F^{\mu \nu}$ is part of a larger field $\psi$ with spin $(1/2,1/2),$ i.e. $\psi$ includes in addition 4-vector $v^{\mu}$ components. This, in spite of the fact that the 4-vector components $v^{\mu}$ do not take part in the field equations. The situation reflects the fact that the Lorentz rep of spin $(1/2,1/2)\oplus(0,1)\oplus(1,0)$ has vector matrices $V^{\mu}$ whereas a Lorentz rep of spin $(0,1)\oplus(1,0)$ does not have vector matrices.

When $a \neq$ 0, the vector matrices $V^{\mu}$ can be rescaled by a factor of $m/a$ and the field equation (\ref{Field6}) becomes
\begin{equation} \label{Field11}  i V^{\mu} \partial_{\mu} \psi =  m  \psi \quad.
 \end{equation}
Thus we have found an equation that has the known field $\psi$ as one solution.

\pagebreak

\appendix

\section{Derivation of Block Scale Factors}

Just one of the four cases is considered. The other cases are similar. We derive $N^{(AB;CD)}_{(1)}$ = $N^{(AB;CD)}_{(1u)},$ rewriting the parameters $r_{u}$ and $s_{u}$ as $r_{u}$ = $r$ and $s_{u}$ = s. The results for all four cases appear in Table 1. 

The parameter $N^{(AB;CD)}_{(1)}$ is the arbitrary scale factor for the block $V^{t}_{(AB,CD;1)}$ for case 1 in (\ref{Cases1}), i.e. $C$ = $A-1/2$ and $D$ = $B-1/2.$ By (\ref{CG2}), (\ref{blockFORM1V}) and (\ref{Field1}), we require the block to be scaled so that
\begin{equation} \label{Vtcg1}  r V^{t}_{(AB,CD;1)} \langle Cc;Dd\mid j j \rangle  = s \langle Aa;Bb\mid j j \rangle \quad.
 \end{equation}
Note that the spin component $\sigma$ is chosen to be $j$, its maximum value. We choose $\sigma$ = $j$ because a fairly simple formula is well known for the Clebsch-Gordon coefficients $\langle j_{1}m_{1};j_{2}m_{2}\mid j j \rangle.$  Altering the notation a little, the formula in Ref. \cite{Messiah1} is as follows,
$$ \langle j_{1}m_{1};j_{2}m_{2}\mid j j \rangle =   \hspace{12cm} $$
\begin{equation} \label{CG1a}  (-1)^{j_{1}-m_{1}}\left( \frac{(2j+1)!(j_{1}+j_{2}-j)!(j_{1}+m_{1})!(j_{2}+m_{2})!}{(j_{1}+j_{2}+j+1)!(j+j_{1}-j_{2})!(j+j_{2}-j_{1})!(j_{1}-m_{1})!(j_{2}-m_{2})!} \right)^{1/2} \quad,
 \end{equation}
with $m_{1}+m_{2}$ = $j.$

A formula for $V^{t}_{(AB,CD;1)}$ can be found by combining the $V^{z} \pm V^{t}$ formulas in Ref. \cite{Beta4},
$$ (V^{t}_{(AB,CD;1)})_{ab;cd} =  \hspace{12cm} $$
\begin{equation} \label{Vt1}  \frac{-N^{(AB;CD)}_{(1)}}{2\sqrt{AB}}\left( \sqrt{(A+a)(B-b)} \delta_{c,a-1/2} \delta_{d,b+1/2} - \sqrt{(A-a)(B+b)} \delta_{c,a-1/2} \delta_{d,b+1/2} \right) \quad,
 \end{equation}
where the factor $N^{(AB;CD)}_{(1)}$  is called $t^{12}_{AB}$ in Ref. \cite{Beta4}. 

Substituting (\ref{CG1a}) and (\ref{Vt1}) in (\ref{Vtcg1}) determines the factor $N^{(AB;CD)}_{(1)},$
\begin{equation} \label{c1}  N^{(AB;CD)}_{(1)} = \frac{2s\sqrt{AB}}{r\sqrt{(A+B)(A+B+1)-j(j+1)}} \quad.
 \end{equation}
The result is listed in Table 1. The other formulas in Table 1 may be calculated in much the same way. 

%\pagebreak

\section{Exercises and Problems} \label{Pb}

\vspace{0.3cm}
\noindent 1. Consider a field $\psi^{(+)},$ (\ref{psiua1}), transforming with a Lorentz irrep with spin $(A,B),$ built with spin $j$ operators ${\mathbf{a}}_{\sigma}.$ Assume the compatibility of spin $j$ with $(A,B),$ (\ref{ABCDj}). Show that when $AB \neq$ 0, the number of components of $\psi^{(+)}$ exceeds the number of components of ${\mathbf{a}}_{\sigma}.$ [Such a field must have field equations because the field cannot have more independent components than there are independent operators. Also note that the fields in the paper all transform with {\it{reducible}} Lorentz reps.]

\vspace{0.3cm}
\noindent 2. Find the number of components in a field $\psi$ that transforms by a Lorentz rep  with spin $(A,B)\oplus$ $ \sum_{(n)} (C_{(n)},D_{(n)}),$ when all the values of $C_{(n)}$ and $D_{(n)}$ in (\ref{Cases1}) are nonnegative. Write the answer as a function of $A$ and $B.$  Answer: $5(2A+1)(2B+1)$

\vspace{0.3cm}
\noindent 3. Rewrite the expressions for the parameters $N^{(AB;CD)}_{(n)}$ and $N^{(CD;AB)}_{(n)}$ in terms of the spins $(C_{(n)},D_{(n)}).$ How is $N^{(CD;AB)}_{(4)}$ related to $N^{(AB;CD)}_{(1)}$? Are there any other, similar relationships?

\vspace{0.3cm}
\noindent 4. Derive formulas for the parameters $N^{(AB;CD)}_{(n)}$ and $N^{(CD;AB)}_{(n)}$ for the cases not derived in the Appendix. Do this by following the steps in the Appendix or otherwise.

\vspace{0.3cm}
\noindent 5. Find the field equations (\ref{Field4}) when the given Lorentz irrep has spin $(A,B)$ = $(0,0)$ and $\psi$ is the field of a scalar particle $j$ = 0. [Hint: One may wish to find a similarity transformation to take the angular momentum and vector matrices from the standard formulas in Ref. \cite{Beta4} to more familiar values with the $(C,D)$ = $(1/2,1/2)$ part of the field $\psi$ transforming recognizably as a 4-vector $v^{\mu}$. Do not confuse $v^{\mu}$ with a coefficient function.]

\vspace{0.3cm}
\noindent 6. Show that the field equations (\ref{Field9}) for a field $\psi$ transforming by the Lorentz rep with spin $(1/2,1/2)\oplus(0,1)\oplus(1,0)$ are the homogeneous Maxwell equations, $F^{[\lambda \mu, \nu]}$ = 0, where $F^{\lambda \mu}$ is an antisymmetric tensor, comma denotes partial differentiation and the brackets indicate the cyclic sum over $\{\lambda, \mu, \nu\}.$  [Hint: One may wish to find a similarity transformation to take the angular momentum and vector matrices from the standard formulas in Ref. \cite{Beta4} to more familiar matrices with the $(1/2,1/2)$ part of the field $\psi$ transforming recognizably as a 4-vector and the $(0,1)$ and $(1,0)$ fields $F^{\lambda \mu}$ transforming under rotations as 3-vector combinations $\overrightarrow{E} + i \overrightarrow{B}$ and $\overrightarrow{E} - i \overrightarrow{B},$ where $B^{i}$ = $\epsilon_{ijk} F^{jk} $ and $E^{i}$ = $F^{it}.$ Thus the transformed $\psi$ has components $\{v^{\mu},\overrightarrow{E} + i \overrightarrow{B},\overrightarrow{E} - i \overrightarrow{B}\}.$] 

%\pagebreak

\end{document}